\documentclass[aps,prb,twocolumn,preprintnumbers,amsmath,amssymb,superscriptaddress,showpacs]{revtex4-1}
\pdfoutput=1
\usepackage{graphicx}
\usepackage{bm}
\usepackage{color}
\usepackage{hyperref}

\begin{document}
\title{Temperature dependence of electronic transport through molecular magnets in the Kondo regime}

\author{Maciej Misiorny}
 \email{misiorny@amu.edu.pl}
\affiliation{Peter Gr\"{u}nberg Institut, Forschungszentrum J\"{u}lich, D-52425 J\"{u}lich, Germany}
\affiliation{JARA -- Fundamentals of Future Information Technologies}
\affiliation{Faculty of Physics, Adam Mickiewicz University,
61-614 Pozna\'{n}, Poland}

\author{Ireneusz Weymann}
\affiliation{Faculty of Physics, Adam Mickiewicz University,
61-614 Pozna\'{n}, Poland}

\author{J\'{o}zef Barna\'{s}}
\affiliation{Faculty of Physics, Adam
Mickiewicz University, 61-614 Pozna\'{n}, Poland}
\affiliation{Institute of Molecular Physics, Polish Academy of
Sciences, 60-179 Pozna\'{n}, Poland
}%


\begin{abstract}
The effects of finite temperature in transport through
nanoscopic systems exhibiting uniaxial magnetic anisotropy $D$,
such as molecular magnets, adatoms, or quantum dots side-coupled to a large spin are analyzed in the Kondo
regime. The linear-response conductance is calculated by means of
the full density-matrix numerical renormalization group method as a function of
temperature $T$, magnetic anisotropy $D$, and exchange coupling $J$ between the molecule's
core spin and the orbital level. It is shown that such system
displays a two-stage Kondo effect as a function of temperature
and a quantum phase transition as a function of the exchange coupling $J$.
Moreover, additional peaks are found in the
linear conductance for temperatures of the order of $T\sim |J|$
and $T\sim D$. It is also shown that the conductance variation
with $T$ remarkably depends on the sign of the exchange coupling $J$.
\end{abstract}

\pacs{73.23.-b,75.50.Xx,85.75.-d,72.15.Qm}


\maketitle


\section{Introduction}

Recent progress in experimental techniques that allow for dealing
with systems involving single atoms or molecules has opened a path
for a new generation of electronic and spintronic
devices.~\cite{Heath_Ann.Rev.Mater.Res.39/2009,McCreery_Adv.Mater.21/2009}
Functionality of such systems is usually based on their magnetic
properties.~\cite{Bogani_NatureMater.7/2008,
Rogez_Adv.Mater.21/2009,Misiorny_Phys.Rev.B79/2009,Misiorny_Europhys.Lett.89/2010}
In particular, the combination of a large spin and uniaxial
magnetic anisotropy makes magnetic
adatoms~\cite{Brune_Surf.Sci.603/2009} and single-molecule magnets
(SMMs)~\cite{Gatteschi_book} promising candidates as
information-storage media.~\cite{Mannini_NatureMater.8/2009,Loth_NaturePhys.6/2010}

The understanding of transport properties of atomic and/or
molecular systems exhibiting magnetic anisotropy in the whole
range of the coupling to electrodes lies at the bottom of their
potential applications. Especially interesting in this context
seems to be the limit of  \emph{strong} coupling, in which some
nontrivial many-body effects stemming from the interplay of
magnetic anisotropy and the Kondo effect are
expected.~\cite{Romeike_Phys.Rev.Lett.96/2006,Romeike_Phys.Rev.Lett.97/2006,Romeike_Phys.Rev.Lett.106/2011,
Leuenberger_Phys.Rev.Lett.97/2006,Misiorny_Phys.Rev.Lett.106/2011,Misiorny_Phys.Rev.B84/2011}
In particular, it turned out that the cooperation of quantum
tunneling and spin-exchange processes may lead to the
pseudo-spin-1/2 Kondo
effect.~\cite{Romeike_Phys.Rev.Lett.96/2006,Romeike_Phys.Rev.Lett.97/2006,Romeike_Phys.Rev.Lett.106/2011}
Moreover, as long as a moderate external magnetic field is
involved, there are no qualitative differences between the
mechanisms of the Kondo effect in systems with  half- and
full-integer spins.~\cite{Leuenberger_Phys.Rev.Lett.97/2006} It
has also been suggested that the formation of the Kondo resonance
should depend on how the system's total spin is modified, i.e.
reduced or augmented, upon accepting a surplus
electron.~\cite{Gonzalez_Phys.Rev.B78/2008} Actually, the Kondo
effect can occur only in the former case, which corresponds to the
antiferromagnetic coupling in the effective spin-1/2 anisotropic
Kondo Hamiltonian. Interestingly enough, by changing the magnitude
of transverse magnetic field one can induce the oscillations of
the Kondo effect, which  stem from the Berry-phase periodical
modulation of the tunnel
splitting.~\cite{Leuenberger_Phys.Rev.Lett.97/2006} Finally, when
the attached electrodes are ferromagnetic, behavior of the tunnel
magnetoresistance (TMR) in the Kondo regime for systems under
discussion is expected to be significantly
different~\cite{Misiorny_Phys.Rev.Lett.106/2011,Misiorny_Phys.Rev.B84/2011}
from that for typical magnetically isotropic quantum
dots.~\cite{Weymann_PhysRevB2005,Barnas_JPCM2008,Weymann_Phys.Rev.B83/2011}

Although the observation of the Kondo-related features in systems displaying magnetic anisotropy
is experimentally challenging as it requires cooling the system down to very low temperatures,
several successful measurements have been recently reported.~\cite{Otte_NaturePhys.4/2008,
Ternes_J.Phys.:Condens.Matter21/2009,Parks_Science328/2010}
Generally, variation of temperature in a nanoscopic system
revealing the Kondo correlations results in a dramatic change of
its transport properties. In quantum dots, lowering the
temperature $T$ below a characteristic energy scale -- the
\emph{Kondo temperature} $T_\textrm{K}$ -- is accompanied by
an increase of the conductance  to its maximum value, which, on
the other hand, becomes gradually suppressed with increasing
$T$.~\cite{Hewson_book,
Goldhaber-Gordon_Nature391/1998,Cronenwett_Science1998,vanderwiel_Science2000}
In the Kondo regime, the dependence of the linear conductance on
temperature is then a universal function of
$T/T_\textrm{K}$.~\cite{Goldhaber-Gordon_Nature391/1998,Cronenwett_Science1998,vanderwiel_Science2000}
Moreover, it was shown very recently that the conductance of
quantum dots coupled to ferromagnetic leads or subject to an
external magnetic field also exhibits universal features with
respect to the (effective) magnetic
field.~\cite{Gaass_Phys.Rev.Lett.107/2011,Kretinin_Phys.Rev.B84/2011}

The temperature dependence of the Kondo effect becomes more
complex when the system consists of more impurity spins, and the
competition between the spin-exchange processes due to tunneling
of electrons and the interaction between the constituent spins is
possible. Even in the conceptually simplest case involving two
exchange coupled spin-1/2 impurities, e.g. as in quantum dots
containing an even number of
electrons,~\cite{Sasaki_Nature405/2000,Tarucha_Phys.Rev.Lett.84/2000}
there are several different scenarios regarding the inter-impurity
exchange interaction $J$.~\cite{Pustilnik_Phys.Rev.Lett.87/2001,
Pustilnik_Phys.Rev.B68/2003,Hofstetter_Phys.Rev.B69/2004,
Pustilnik_Phys.Rev.B73/2006,
Logan_Phys.Rev.B80/2009,Florens_J.Phys.:Condens.Matter23/2011} In
the low temperature limit, the impurities form a singlet ($S=0$)
for  large \emph{antiferromagnetic} $J$ and the effect of
conduction electrons on the system is weak, while the high-spin
triplet ($S=1$) ground state  develops for the
\emph{ferromagnetic} coupling, which then can be screened. A
similar scenario is also relevant for side-coupled double quantum
dots in single spin regime, with only one dot coupled directly to
conduction electrons. In such a setup, depending on the strength
and sign of the spin exchange interaction between the two dots, a
two-stage Kondo effect and Kosterlitz-Thouless quantum phase
transition can occur.~\cite{Kusunose_J.Phys.Soc.Jpn.66/1997,
Vojta_Phys.Rev.B65/2002,Cornaglia_Phys.Rev.B71/2005,
Chung_PhysRevB.77.035120,Zitko_J.Phys.:Condens.Matter22/2010}

As the magnitude of the impurity spin becomes larger than 1/2, a
new energy scale related to magnetic anisotropy enters the
problem. In the case of a single \emph{anisotropic} Kondo impurity
with spin $S>1/2$ coupled to a single conduction-electron channel,
which accurately represents the situation of a magnetic adatom
deposited on a nonmagnetic surface,~\cite{Loth_NaturePhys.6/2010,
Otte_NaturePhys.4/2008,Ternes_J.Phys.:Condens.Matter21/2009} it
has been shown that at low temperatures the spin is ultimately
always subject to complete
compensation.~\cite{Zitko_Phys.Rev.B78/2008,Zitko_NewJ.Phys.11/2009}
In this paper we consider more complex systems exhibiting an
uniaxial magnetic anisotropy, e.g. adatoms, quantum dots
side-coupled to a large spin $S$, as well as molecular magnets, where
additionally the charge state of the system can be changed owing
to electron tunneling processes. The central aim is then to study
the finite-temperature transport properties of such systems,
focusing on the Kondo regime. Understanding the behavior of system
at different temperatures is of great importance, as many
experiments are actually carried out in the cross-over regime,
$T\sim T_K$, when neither Fermi liquid ($T\ll T_K$) nor
perturbative ($T\gg T_K$) descriptions are applicable. Our
analysis is based on the full density-matrix numerical
renormalization group (NRG)
method,~\cite{Wilson_Rev.Mod.Phys.47/1975,Bulla_Rev.Mod.Phys.80/2008,
Weichselbaum_Phys.Rev.Lett.99/2007} which allows for calculating
transport at any temperature in an essentially exact way.

\section{Theoretical description}

\begin{figure}[t]
  \includegraphics[width=0.65\columnwidth]{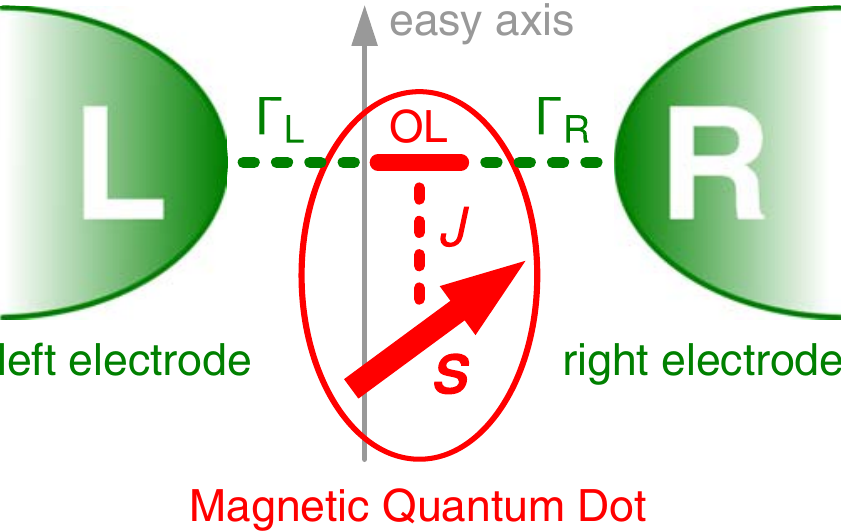}
  \caption{\label{Fig:system} (Color online)
  Schematic depiction of the system under consideration.
  It consists of a single orbital level (OL) tunnel-coupled to external leads,
  with coupling strengths $\Gamma_L$ and $\Gamma_R$ for the left and
  right leads, and additionally exchanged-coupled to a spin $S$,
  with $J$ denoting the strength of exchange interaction.}
\end{figure}

The key features of systems under consideration -- such as
magnetic adatoms, quantum dots coupled to localized magnetic
impurities, and SMMs --  can be reproduced by a model consisting
of a single conducting orbital level (OL) exchange-coupled with strength $J$ to a
magnetic core (spin $S$) subject to uniaxial magnetic anisotropy $D$, see Fig.~\ref{Fig:system}.
This system will be referred to as \emph{magnetic quantum dot} (MQD) and its
Hamiltonian has the form~\cite{Romeike_Phys.Rev.Lett.96/2006,
*Romeike_Phys.Rev.Lett.97/2006,Romeike_Phys.Rev.Lett.106/2011,
*Leuenberger_Phys.Rev.Lett.97/2006,Misiorny_Phys.Rev.Lett.106/2011}
    \begin{equation}\label{Eq:1}
    \mathcal{H}_\textrm{MQD}=\mathcal{H}_\textrm{OL} - D S_z^2 -J \bm{s}\cdot\bm{S}.
    \end{equation}
The orbital level is described by $\mathcal{H}_\textrm{OL}=\varepsilon\sum_{\sigma}  n_\sigma
+ U n_\uparrow n_\downarrow$, where $n_\sigma=c_\sigma^\dagger c_\sigma^{}$ is the occupation
operator, with $c_\sigma^\dagger (c_\sigma^{})$
creating (annihilating) a spin-$\sigma$ electron of energy $\varepsilon$ in the orbital level,
while $U$ accounts for the Coulomb energy of two electrons of opposite spins
residing in the orbital. Furthermore, the second term of the Hamiltonian~(\ref{Eq:1})
characterizes the magnetic anisotropy of the core, where $D$ denotes the
uniaxial magnetic anisotropy constant, and $S_z$ is the $z$th component of the
MQD's internal spin operator $\bm{S}$. The present discussion is limited only
to the case of systems exhibiting magnetic bistability ($D>0$).
Finally, the exchange interaction between the magnetic core of a
MQD and the spin of an electron occupying the orbital level, given by
$\textbf{s}=\frac{1}{2}\sum_{\sigma\sigma'}c_\sigma^\dag
\bm{\sigma}_{\sigma\sigma'}^{} c_{\sigma'}^{}$,
with $\bm{\sigma}\equiv(\sigma^x,\sigma^y,\sigma^z)$ standing for the Pauli spin
operator, is expressed by the last term of $\mathcal{H}_\textrm{MQD}$.
In general, the $J$-coupling can be either of \emph{ferromagnetic}
(FM for $J>0$) or \emph{antiferromagnetic} (AFM for $J<0$) type.

The magnetic quantum dot is assumed to be tunnel-coupled to two identical
electrodes only \emph{via} the orbital level, see Fig.~\ref{Fig:system}, and electrons in the $q$th electrode
[$q=\ $(\emph{L})eft, (\emph{R})ight] are modelled by,
$\mathcal{H}_\textrm{el}^q=\sum_{\sigma}\int_{-W}^W\textrm{d}\epsilon\,
\epsilon \, a_{q\sigma}^{\dagger}(\epsilon)a_{q\sigma}(\epsilon)$, with
$a_{q\sigma}^{\dagger}(\epsilon)$ being the relevant creation
operator, and $W$ the band half-width.
The electron tunneling processes between the MQD and electrodes are
described by
\begin{equation}
  \mathcal{H}_\textrm{tun}=\sum_{q\sigma}\sqrt{\frac{\Gamma_q}{\pi}}
  \int_{-W}^W\textrm{d}\epsilon\, [a_{q\sigma}^{\dagger}(\epsilon) c_\sigma^{}
  + c_\sigma^\dagger a_{q\sigma}(\epsilon)],
\end{equation}
where $\Gamma_q$ represents the strength of coupling of the orbital level to the $q$th lead.

Conceptually, the model considered is equivalent to a single-level
quantum dot which -- if occupied by a single electron -- is
exchanged coupled to a large-spin magnetic impurity subject to
magnetic anisotropy.~\cite{Kusunose_J.Phys.Soc.Jpn.66/1997}
Moreover, to some extent it can be regarded as an alternative to a
two-impurity Kondo model where only one impurity couples directly
to conduction band, or to a double quantum dot system in a T-shape
geometry.~\cite{Vojta_Phys.Rev.B65/2002,Cornaglia_Phys.Rev.B71/2005,
Chung_PhysRevB.77.035120,Zitko_J.Phys.:Condens.Matter22/2010} From
this point of view our model will also exhibit a two-stage Kondo
effect and a quantum phase transition, as shall be discussed in
next section.

Generally, the physics of the Kondo effect is essentially determined
by the number of conduction-electron channels to which the system is coupled,
as in order to completely screen a spin $S$ one needs $2S$ channels.~\cite{Nozieres_J.Physique41/1980}
In this context, high-spin molecular devices are unique as they commonly
operate in the regime where effectively only one
channel plays a role.~\cite{Parks_Science328/2010,Florens_J.Phys.:Condens.Matter23/2011}
More precisely, even if the device is coupled to multiple leads,
each of such junction usually supports only a single conduction channel,
and additionally the couplings are typically characterized by a strong asymmetry.
As a result, the relevant Kondo energy scale is determined by the strongest coupling,
since the Kondo temperatures associated with weaker
couplings are exponentially small,~\cite{Haldane_Phys.Rev.Lett.40/1978}
and hence negligible under typical experimental conditions.
In the model considered here the situation is even simpler, since
the orbital level of magnetic quantum dot couples only to an even linear combination
of the electron operators in the left and right leads, with a new coupling strength
$\Gamma \equiv \Gamma_L+\Gamma_R$, while the odd combination is completely decoupled.
This basically means that we can limit our discussion to the case of a \emph{single} conduction-electron channel.

In the following, we will study the linear-response transport properties of magnetic quantum dots
on various parameters of the system. The main quantity we are interested in is the
temperature-dependent  \emph{linear conductance} $G$, which we
calculate from the Landauer-Wingreen-Meir
formula,~\cite{Landauer_Philos.Mag.21/1970,Meir_Phys.Rev.Lett.68/1992}
    \begin{equation}
    G=\frac{2e^2}{h}\sum_\sigma\frac{2\Gamma_L\Gamma_R}
    {\Gamma_L+\Gamma_R}\int \! \textrm{d}\omega
    \left(\! -\frac{\partial f(\omega)}{\partial \omega}\right)\pi
    A_\sigma(\omega),
    \end{equation}
where $f(\omega)$ denotes the Fermi-Dirac distribution function,
while $A_\sigma(\omega)$ is the spin-dependent \emph{spectral
function} of the orbital level, and $2\Gamma_L\Gamma_R/(\Gamma_L+\Gamma_R)=\Gamma/2$ in the present situation.
The problem of determining the MQD's transport features corresponds then essentially
to finding the spin-resolved spectral function $A_\sigma(\omega)$, which in the
present work is obtained by means of the Wilson's numerical
renormalization group (NRG) method.~\cite{Wilson_Rev.Mod.Phys.47/1975,Bulla_Rev.Mod.Phys.80/2008}
In particular, we employ the recent idea of a full density
matrix,~\cite{Weichselbaum_Phys.Rev.Lett.99/2007} which allows for
reliable calculation of static and dynamic properties of the
system at arbitrary temperatures.~\cite{Legeza_DMNRGmanual} For
the present problem, to obtain decent results, the $U_\textrm{charge}(1)\times
U_\textrm{spin}(1)$ symmetry was exploited, the discretization
parameter $\Lambda=1.8$ was used and we kept $N_k=1200$ states
during calculations.

\section{Results and discussion}

In this paper we consider the transport features of a
prototypical magnetic quantum dot characterized by the spin $S=2$.
The other parameters are typical of molecular systems, see the caption of
Fig.~\ref{Fig:2}. In order to discuss the influence of finite
temperature on the Kondo effect, we introduce the Kondo
temperature $T_\textrm{K}$, to which other parameters will be
compared whenever it is useful. The Kondo temperature is defined
here as the half-width at the half-maximum of the
normalized linear conductance $G/G_{T,J,D=0}$
as a function of $T$, where $G_{T,J,D=0}$ is the
conductance for $T=D=J=0$, which yields
$T_\textrm{K}/U\approx5\cdot10^{-3}$.

\subsection{The case of  \emph{zero} magnetic anisotropy ($D=0$)}

\begin{figure}[t]
  \includegraphics[width=0.95\columnwidth]{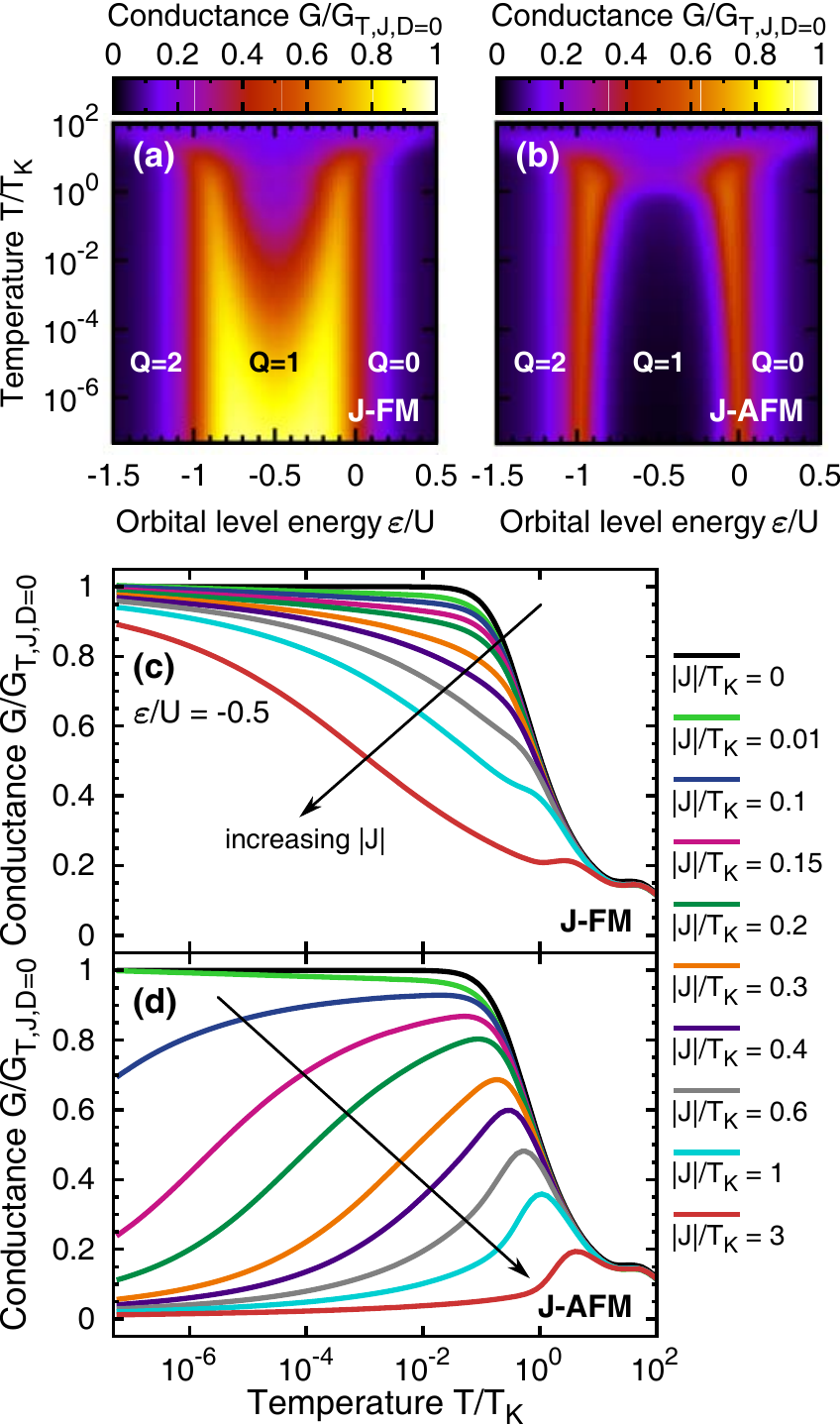}
  \caption{\label{Fig:2} (Color online)
  The dependence of the linear conductance $G$ on the energy
  of the orbital level (OL) $\varepsilon$ and the temperature $T$
  in the case of (a,c) \emph{ferromagnetic} (FM) and (b,d)
  \emph{antiferromagnetic} (AFM) exchange coupling $J$ for $|J|/T_\textrm{K}=3$ and $D=0$.
  The variable $Q$ represents the average number of
  electrons that occupy the OL. Here,  $G_{T,J,D=0}$
  corresponds to the conductance calculated for $T=D=J=0$.
  Parts (c)-(d) present the relevant cross-sections
  of the density plots  at the electron-hole symmetry
  point ($\varepsilon=-U/2$) for different values of the $J$-coupling, as indicated.
  The parameters of the system are as follows:
  $U=10$ meV, $\Gamma/U=0.1$ and $T_\textrm{K}/U\approx5\cdot10^{-3}$.
  }
\end{figure}

Before the influence of the magnetic anisotropy on the temperature
dependence of the Kondo effect is analyzed, it is instructive to
consider the situation with vanishing magnetic anisotropy, $D=0$.
The corresponding results are shown in Fig.~\ref{Fig:2}, which
presents the dependence of the linear conductance on orbital level
position and temperature for different exchange couplings $J$. It
can be seen that when the orbital level is occupied by even number
of electrons, which corresponds to an empty or fully occupied
level, the conductance is generally suppressed and determined only
by elastic cotunneling processes, see Figs.~\ref{Fig:2}(a,b). On
the other hand, for an odd occupation of the orbital level (single
spin regime), the Kondo effect should in general develop at low
temperatures. Now, however, the exchange interaction with the core
spin $S$ comes into play. In principle, two distinctive cases with
respect to the $J$-coupling sign can be recognized. For the
\emph{ferromagnetic} exchange coupling, the low temperature
behavior of the system is governed by the competition between the
coupling $J$ and hybridization with electrodes $\Gamma$. The
former interaction tends to stabilize the high-spin state, whereas
the later one leads to the screening of the orbital level's spin.
It has been shown for a system including two spin-1/2 impurities,
that the Kondo effect dominates even if the exchange coupling is
significantly larger than the
hybridization.~\cite{Kusunose_J.Phys.Soc.Jpn.66/1997,Izumida_J.Phys.Soc.Jpn.67/1998}
This can be also seen in Fig.~\ref{Fig:2}(c), where irrespective
of the magnitude of $J$, $G$ tends to its maximum value in the
zero temperature limit. Nevertheless, the exchange interaction
still plays a prominent role, because it is responsible for the
reduction of the Kondo temperature, as compared to a simple
spin-1/2 magnetic impurity system.

The situation is qualitatively different for
\emph{antiferromagnetic} $J$-coupling. Investigations of high-spin
two-impurity (with spins $S_1$ and $S_2$) Kondo models
characterized by the \emph{antiferromagnetic} inter-impurity
exchange interaction and only one impurity ($S_1$) directly
coupled to a conduction band have revealed a two-stage Kondo
effect which is a generic feature for all $S_2\geqslant S_1$
models.~\cite{Zitko_J.Phys.:Condens.Matter22/2010} In particular,
a two-stage Kondo process should take place for the inter-impurity
exchange coupling smaller than the energy scale
$T_\textrm{K}$ associated with the Kondo screening of the
directly coupled impurity. For $S_1=1/2$, the temperature
$T_\textrm{K}^\prime$ at which $S_2$ becomes screened is then
exponentially smaller, as the process occurs due to interaction of
$S_1$ with a Fermi sea arising as a consequence of the first
screening stage.~\cite{Cornaglia_Phys.Rev.B71/2005} The two-stage
Kondo effect can be nicely seen in Fig.~\ref{Fig:2}(d), which
shows the temperature dependence of linear conductance for
different \emph{antiferromagnetic} exchange coupling. Let us have
a closer look at the curve corresponding to $|J|/T_K = 0.2$. For
$T<T_K$, the conductance starts increasing due to the Kondo effect
associated with the screening of the orbital level spin by
conduction electrons. However, at sufficiently low temperatures
the energy scale related with the second stage of screening
becomes relevant and the conductance becomes suppressed. Moreover,
when varying the strength of the exchange interaction at zero
temperature, the system undergoes a quantum phase transition.
This phase transition is similar to a singlet-triplet transition
observed in multilevel quantum dots,~\cite{Hofstetter_Phys.Rev.Lett.88/2001,
Kogan_Phys.Rev.B67/2003,Hofstetter_Phys.Rev.B69/2004,
Pustilnik_Phys.Rev.B73/2006,Roch_Nature453/2008,Florens_J.Phys.:Condens.Matter23/2011}
and for finite temperature, magnetic field or anisotropy turns into a crossover.

\begin{figure}[t]
  \includegraphics[width=0.95\columnwidth]{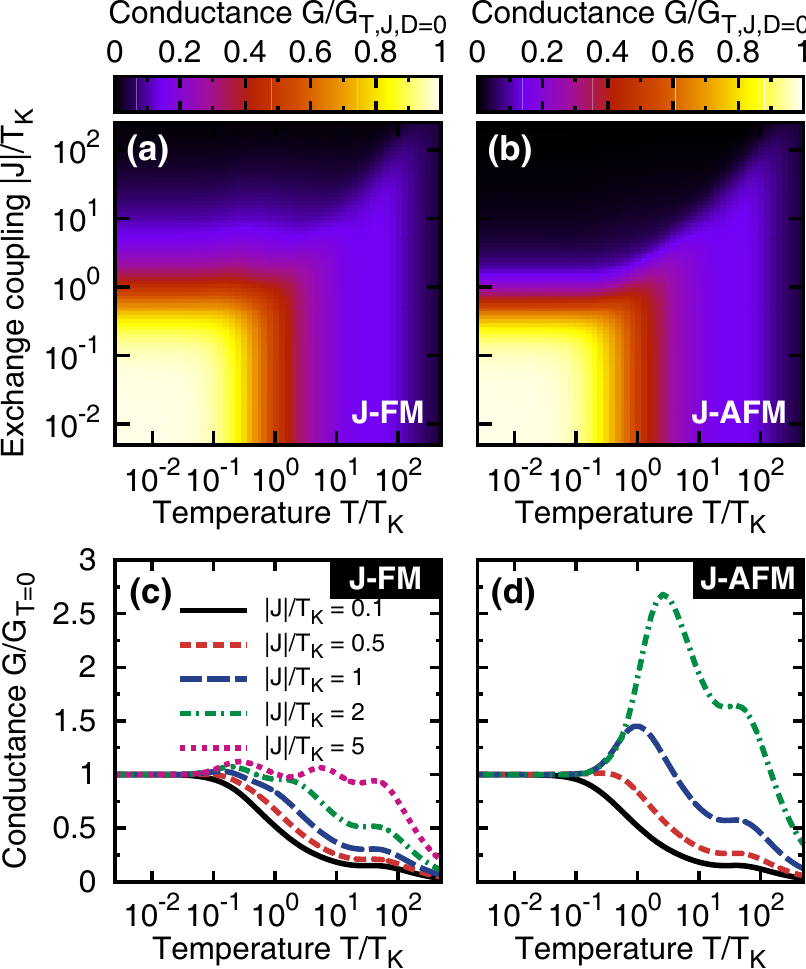}
  \caption{\label{Fig:3} (Color online)
  The normalized linear conductance $G$ as a function of temperature $T$
  and exchange coupling parameter $J$ in the case of (a,c)
  \emph{ferromagnetic} (FM) and (b,d) \emph{antiferromagnetic} (AFM) type of the
  coupling between magnetic core and orbital level shown in the presence of uniaxial magnetic anisotropy,
  $D/U=10^{-4}$ ($D/T_\textrm{K}=0.2$).
  Curves in (c)-(d) represent cross-sections of figures (a) and (b), respectively,
  for indicated values of $|J|$, but now each curve is normalized
  to the corresponding $G_{T=0}$ instead of $G_{T,J,D=0}$ as in (a)-(b).
  The other parameters are the same as in Fig.~\ref{Fig:2} with $\varepsilon=-U/2$.}
\end{figure}

\subsection{The case of \emph{finite} magnetic anisotropy ($D>0$)}

\begin{figure*}[t]
  \includegraphics[width=0.95\textwidth]{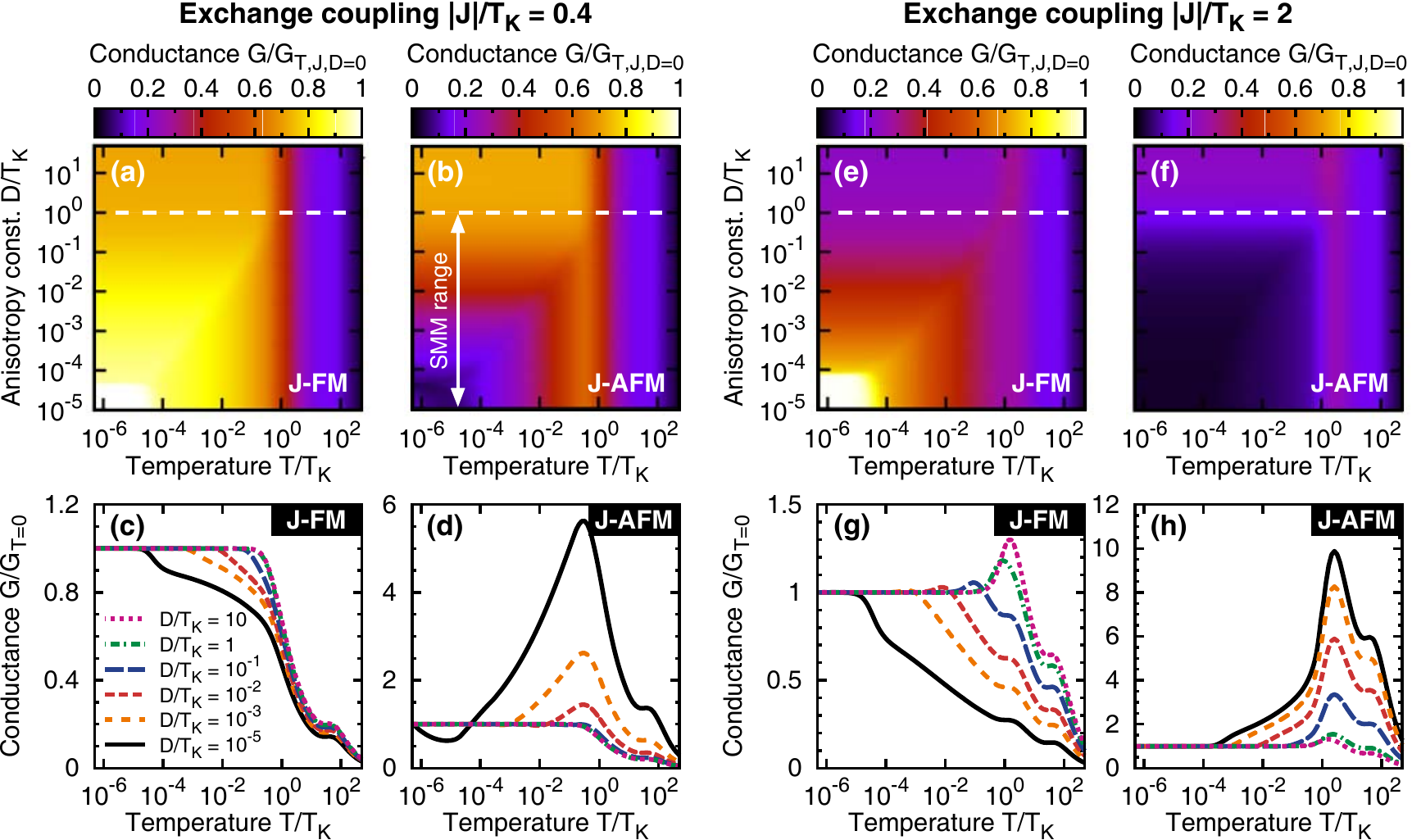}
  \caption{\label{Fig:4} (Color online)
  The normalized linear conductance $G$ as a function
  of temperature $T$ and uniaxial magnetic anisotropy $D$
  for $|J|/ T_\textrm{K}=0.4$ in (a)-(d) and $|J|/  T_\textrm{K}=2$ in (e)-(h).
  The other parameters are the same as in Fig.~\ref{Fig:2} with $\varepsilon=-U/2$.
  Dashed lines in (a)-(b) and (e)-(f) represent maximal value of the
  uniaxial magnetic anisotropy constant $D$
  typical to molecular magnets, for further details see the text.}
\end{figure*}

In the case of finite magnetic anisotropy $D$, the situation
becomes more complex since now the degeneracy of spin multiplets
is partially lifted. More precisely, in the absence of magnetic anisotropy $D$,
the ground state of the MQD is the spin multiplet $S+\frac{1}{2}$ ($S-\frac{1}{2}$)
for ferromagnetic (antiferromagnetic) exchange interaction $J$.
In the case of finite magnetic anisotropy ($D>0$), the ground state
becomes two-fold degenerate and consists of the lowest-weight and
highest-weight components of the above multiplets, respectively.~\cite{Misiorny_Phys.Rev.B84/2011}

Figure~\ref{Fig:3} presents the dependence of
the system's linear conductance $G$ on the exchange coupling $J$
and temperature $T$, for the ferromagnetic and antiferromagnetic
types of the $J$-coupling. As long as $|J|\ll T_\textrm{K}$, the
temperature behavior of the conductance does not differ from that
for a typical single-level quantum
dot,~\cite{Wiel_Science289/2000,Kretinin_Phys.Rev.B84/2011,Weymann_Phys.Rev.B83/2011}
see the  parts (a) and (b) in Fig.~\ref{Fig:3}. The conductance
also does not depend on the sign of the parameter $J$, compare the
solid lines (for $|J|/T_K=0.1$) in Fig.~\ref{Fig:3}(c)-(d). In
such a limit, electrons tunneling through the orbital level are
hardly affected by the presence of the MQD's magnetic core.
However, as $|J|\gtrsim T_\textrm{K}$, the screening of the
orbital level's spin by conduction electrons is suppressed due to
the strong exchange interaction of the orbital level's spin with
the magnetic core. One can then observe that while for $T\ll
T_\textrm{K}$ the conductance $G$ just decreases monotonically
with increasing $|J|$, some additional features of $G$ emerge for
$T\sim T_\textrm{K}$, see Figs.~\ref{Fig:3}(c)-(d). This stems
from the fact that for large $|J|$ the influence of the magnetic
core on the orbital level cannot be neglected, as the tunneling
processes actually take place \emph{via} the molecular spin states
formed due to the $J$-coupling.

Since transport properties of the magnetic quantum dot in the
linear response regime depend basically on the system's ground
state, thus for $|J|\gtrsim T_\textrm{K}$, it is the sign of the
exchange parameter $J$ that determines the system's spin
multiplets that plays the dominant role, i.e. $S+1/2$ for
ferromagnetic $J$ and $S-1/2$ for antiferromagnetic $J$.
Specifically, at low temperatures the MQD with ferromagnetic $J$
is found to occupy the doublet state $S_z=\pm5/2$ of the spin
multiplet $S=5/2$, whereas under the same conditions the system
with antiferromagnetic $J$ prefers the state $S_z=\pm3/2$
belonging to the spin multiplet $S=3/2$. Noting this, the
characteristic behavior of the linear conductance for temperatures
around $T_\textrm{K}$ in Fig.~\ref{Fig:3} can be explained
straightforwardly. In general, the conductance displays additional
features (peaks) whenever the number of MQD's states participating
in electronic transport changes due to increased temperature. For
this reason, as soon as $T\sim D$, the neighboring states of the
same spin multiplet become accessible for electron tunneling
processes, and when $T$ reaches the value of $|J|$, $T\sim|J|$,
also the states of the other spin multiplet enter into
consideration. At these temperatures the linear conductance
displays additional peaks, see Figs.~\ref{Fig:3} (c)-(d).
Nevertheless, it should be emphasized that energies of the MQD's
states with the orbital level occupied by a single electron
generally depend both on $J$ and
$D$.~\cite{Misiorny_Phys.Rev.B84/2011}

Additional feature visible in Fig.~\ref{Fig:3} is an asymmetry
between the ferromagnetic and  antiferromagnetic coupling $J$. Note, however, that the
value of $G_{T=0}$, to which the conductance in Figs.~\ref{Fig:3}
(c)-(d) is normalized, is different in both cases, and the
difference is visible only if $|J|\gtrsim T$, otherwise the
thermal fluctuations smear the difference between the cases of ferromagnetic
and antiferromagnetic exchange coupling, Figs.~\ref{Fig:3} (a)-(b). This is
related to different spin multiplets relevant for spin flip
processes that drive the Kondo effect at low
temperatures.~\cite{Misiorny_Phys.Rev.B84/2011}

\begin{figure}[t]
  \includegraphics[width=0.99\columnwidth]{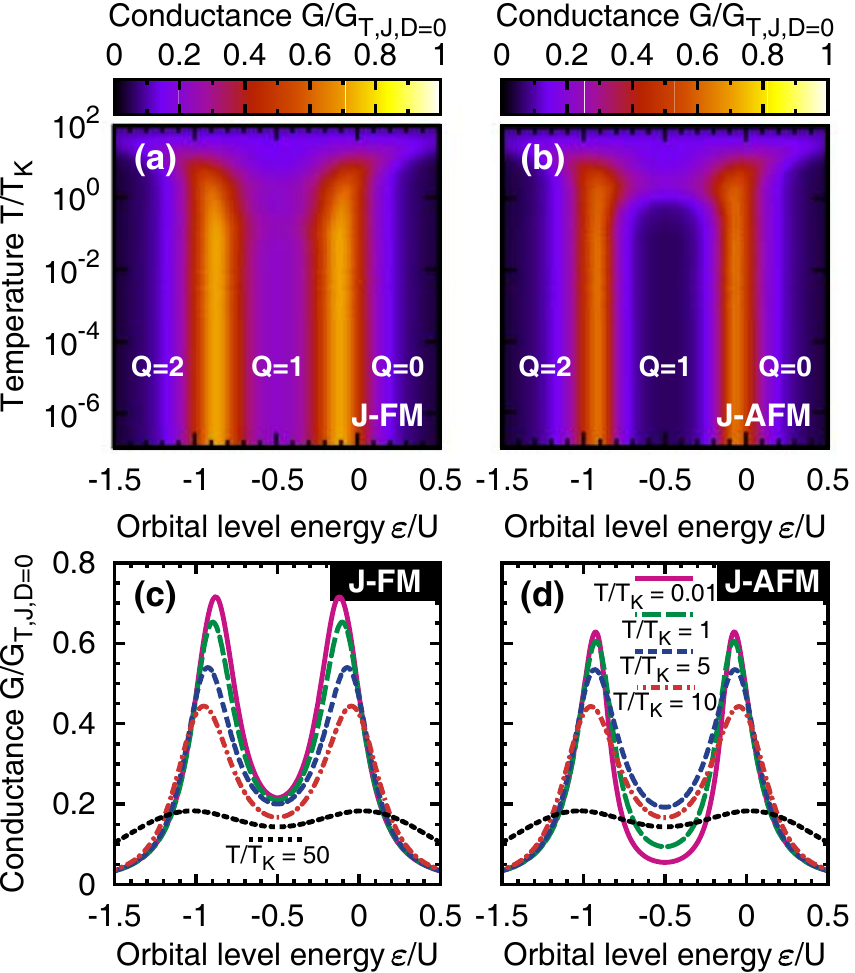}
  \caption{\label{Fig:5} (Color online)
  The normalized linear conductance $G$ as a function of temperature $T$
  and orbital level position $\varepsilon$.
  Parts (c)-(d) present the relevant cross-sections
  of the density plots for different temperatures, as indicated.
  The parameters are the same as in Fig.~\ref{Fig:2} with $|J|/T_K=3$
  and $D/T_K=0.2$.}
\end{figure}

Let us now focus on the effects due to the magnetic anisotropy
$D$, see Fig.~\ref{Fig:4}. It is worth noting that the anisotropy
constant in SMMs can take different values, e.g. in the case of a
Fe$_4$ molecule one finds $D/U\approx5\cdot10^{-3}$. This is
actually one of the highest values of $D$ observed in SMMs. On the
other hand, in magnetic adatoms, such as e.g.
Fe,~\cite{Hirjibehedin_Science317/2007} this constant can be as
large as $D/U\sim0.1$. Therefore, we study the effects resulting
from magnetic anisotropy and finite temperature for a broad range
of anisotropy constant. Figure~\ref{Fig:4} shows the linear
conductance {\em vs.} temperature and anisotropy constant for both
positive and negative exchange coupling $J$. It can be seen that
significant differences between the case of the ferromagnetic and
antiferromagnetic $J$-coupling  occur for $D\ll|J|$. When the
anisotropy is weak, the behavior of linear conductance is the same
as discussed previously in the case of $D=0$, i.e. at low
temperatures the system is in the underscreened Kondo regime for
ferromagnetic $J$-coupling, while for the antiferromagnetic
$J$-coupling one observes a two-stage Kondo effect. When the
anisotropy increases, see e.g. the case of $D = T_K$, both the
underscreened and two-stage Kondo effects become generally
suppressed. Moreover, the temperature dependence of the normalized
linear conductance displays a qualitatively different behavior for
the ferromagnetic and antiferromagnetic type of the $J$-coupling.
While in the case of ferromagnetic $J$ the increase of $T$
generates a drop of the conductance, see Figs.~\ref{Fig:4}(a,c),
the opposite trend appears for antiferromagnetic $J$, see
Figs.~\ref{Fig:4}(b,d). This is mainly related with the fact that
for a given value of $|J|$, the zero-temperature conductance is
much smaller for the antiferromagnetic case than for the
ferromagnetic one. Interestingly enough, in both situations one
can notice a peak at $T\sim|J|$. For this temperature, the
molecular states of another multiplet corresponding to the singly
occupied orbital level become available for transport. Because of
it, the overall rate of spin-flip processes is increased due to
thermal fluctuations and an additional resonance in the
conductance appears. Finally, when $D\gtrsim|J|$, the differences
between $J<0$ and $J>0$ actually disappear, see the dotted curves
for $D/T_K=10$ in Figs.~\ref{Fig:4}(c)-(d), which are almost
identical. Note that the influence of magnetic anisotropy of the
core spin on the transport properties of the system is only
present for considerable exchange couplings $|J|\gtrsim T_K$.

\begin{figure}[t]
  \includegraphics[width=0.95\columnwidth]{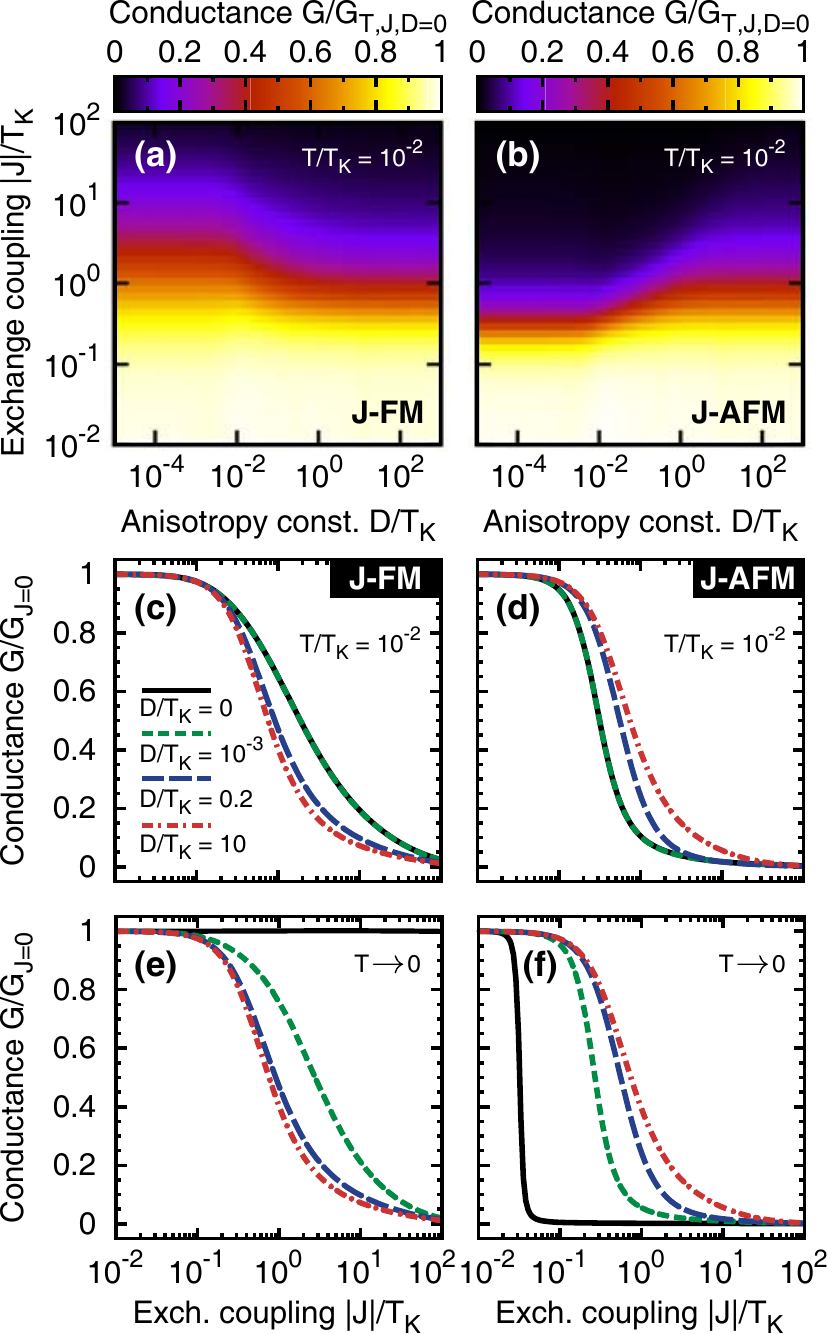}
  \caption{\label{Fig:6} (Color online)
  Dependence of the normalized linear conductance $G$ on the magnetic anisotropy $D$
  and the exchange coupling $J$ of (a) the \emph{ferromagnetic} (FM)
  and (b) \emph{antiferromagnetic} (AFM) type for $T/T_\textrm{K}=0.01$.
  Parts (c) and (d) show the dependence of $G$ on (c) ferromagnetic
  and (d) antiferromagnetic exchange coupling $J$ for $T/T_K = 10^{-2}$
  and for a few anisotropy constants $D$, as indicated.
  Parts (e) and (f) show the same calculated for $T \to 0$.
  The other parameters are the same as in Fig.~\ref{Fig:2} with $\varepsilon=-U/2$.}
\end{figure}

It is also interesting to analyze the temperature dependence of
the linear conductance in the presence of magnetic anisotropy and
for varying position of the orbital level $\varepsilon$, as shown
in Fig.~\ref{Fig:5}. The orbital level position can be
experimentally tuned by a gate voltage. By changing $\varepsilon$,
the orbital level becomes consecutively occupied with electrons,
see Fig.~\ref{Fig:5} where $Q$ denotes the average number of
electrons. At low temperatures, $T\ll \Gamma$, transport for even
occupations is mediated by cotunneling processes and conductance
is low. For the odd occupation ($Q=1$) and $T<T_K$, the Kondo
effect should generally develop. It is however suppressed due to
finite exchange coupling and magnetic anisotropy. It is very
instructive to compare Figs.~\ref{Fig:5}(a)-(b) with
Figs.~\ref{Fig:2}(a)-(b), where actually the same dependence is plotted
for finite anisotropy (Fig.~\ref{Fig:5}) and for $D=0$ (Fig.~\ref{Fig:2}).
It can be seen that the main difference occurs
in the case of ferromagnetic exchange coupling, when finite
anisotropy leads to the suppression of the Kondo effect and
conductance is much lower than in the case of $D=0$. In addition,
when $T\sim|J|$, a resonance due to thermally-activated spin-flip
processes through other spin multiplets occurs. This effect is
more visible in the case of antiferromagnetic $J$. On the other
hand, once $T>|J|$, the conductance starts decreasing and the
difference between the cases of positive and negative $J$ is
diminished.

Finally, we study the  normalized linear conductance as a function
of the anisotropy constant $D$ and exchange coupling $J$ for a
constant temperature, $T/T_K = 10^{-2}$, see
Figs.~\ref{Fig:6}(a)-(b). One can see that the effect of exchange
interaction becomes visible when $|J|\gtrsim T_K$. If this is the
case, then the transport properties also depend on the anisotropy
constant $D$. For a fixed value of $|J|$, with $|J|\gtrsim T_K$,
the conductance becomes suppressed by increasing the anisotropy
constant in the case of ferromagnetic exchange interaction, while
for antiferromagnetic $J$, the behavior is just opposite -- there
is an increase of $G$. This tendency is also nicely seen in
the cross-sections shown in Figs.~\ref{Fig:6}(c)-(d) for a few anisotropy constants $D$.
Generally, the conductance drops with increasing $J$ irrespective of the type of exchange interaction.
This is, however, because the curves present $G$ \emph{vs.} $J$ at finite $T/T_K=10^{-2}$.
In the case of antiferromagnetic exchange coupling
there is a quantum phase transition as $J$ is varied.
This can be seen in Fig.~\ref{Fig:6}(f), which was calculated
for temperature $T \to 0$.
This transition turns into a cross-over in the case of finite $T$ and $D$,
see Figs.~\ref{Fig:6}(d) and (f).
In the case of ferromagnetic exchange interaction in the absence of magnetic anisotropy
and for $T \to 0$, the conductance does not depend on $J$ and
equals $2e^2/h$, see Fig.~\ref{Fig:6}(e).
For finite anisotropy $D$, the degeneracy of the ground state multiplet $S+\frac{1}{2}$
is lifted and the Kondo effect is suppressed once $|J| \gtrsim T_K$.
On the other hand, if the temperature is finite,
then at certain value of ferromagnetic exchange interaction the conductance drops
since the screening temperature depends on $J$ and the condition $T<T_K$ is not met any more.


\section{Summary}

We have studied transport properties of magnetic quantum dots
coupled to external leads in the Kondo regime. The analyzed system
 consisted of a spin $S=2$ exchange coupled to a single
orbital level that was directly tunnel-coupled to electrodes. In
particular, we have focused on the dependence of the
linear-response conductance of the system on temperature, orbital
level position, magnetic anisotropy and exchange coupling. The
calculations were performed with the aid of full density-matrix
numerical renormalization group method. In the absence of magnetic
anisotropy, the model studied generally exhibits a two-stage Kondo
effect and an underscreened Kondo phenomenon, depending on the
sign of the exchange coupling. We have shown that these two
effects become generally suppressed in the presence of magnetic
anisotropy, if the exchange coupling is sufficiently strong,
$|J|\gtrsim T_K$. We have also shown that the temperature
dependence of the conductance depends on the type of exchange
coupling $J$. For ferromagnetic coupling, the linear conductance
was found to generally decrease with $T$, while for
antiferromagnetic coupling, the conductance displayed a maximum
for $T\sim |J|$. In addition, the conductance variation with the
exchange coupling reveals a quantum phase transition for
antiferromagnetic $J$-coupling, which turns
into a crossover in the case of finite temperature and magnetic
anisotropy.

The model studied in the present paper can be used to
describe transport through single-molecule magnets,
adatoms or quantum dots exchange-coupled to a large spin $S$.
Our results and predictions may be thus relevant for a wide class of molecular devices.
Finally, we note that molecular devices are more suitable for
observing Kondo-related effects, since they provide larger energy scales,
which translates into higher Kondo temperatures,
more easily achievable in experiments.~\cite{Florens_J.Phys.:Condens.Matter23/2011}


\begin{acknowledgments}

We acknowledge support from the Polish Ministry of Science and
Higher Education through a Iuventus Plus research project. The authors also acknowledge support from the Foundation for Polish Science (M.M.),
 the Humboldt
Foundation (I.W.) and the 7FP of the EU under REA grant agreement No.
CIG-303 689 (I.W.).

\end{acknowledgments}



%

\end{document}